\documentclass[prl, showpacs, twocolumn,superscriptaddress]{revtex4-1}
\usepackage{amsmath}
\usepackage{graphicx}
\usepackage{subfigure}
\usepackage{hyperref}

\newcommand{\be}{\begin{equation}}
\newcommand{\ee}{\end{equation}}
\newcommand{\rss}{rescaled steady state}
\newcommand{\ws}{\tilde w} 

\begin{document}
\title{The Effect of Growth on Equality in Models of the Economy}

\author{Kang Liu}
\email{kangliu@physics.bu.edu}
\affiliation{Department of Physics, Boston University, Boston, Massachusetts 02215}
\author{N. Lubbers}
\email{nlubbers@buphy.bu.edu}
\affiliation{Department of Physics, Boston University, Boston, Massachusetts 02215}
\author{W. Klein}
\email{klein@bu.edu}
\affiliation{Department of Physics, Boston University, Boston, Massachusetts 02215}
\affiliation{Center for Computational Science, Boston University, Boston, Massachusetts 02215}
\author{J. Tobochnik}
\affiliation{Department of Physics, Boston University, Boston, Massachusetts 02215}
\affiliation{Department of Physics, Kalamazoo College, Kalamazoo, MI, 49006} 
\author{B. Boghosian}
\affiliation{Department of Mathematics, Tufts University, Medford, MA 02155}
\affiliation{American University of Armenia, Yerevan 0019, Republic of Armenia}
\author{Harvey Gould}
\affiliation{Department of Physics, Boston University, Boston, Massachusetts 02215}
\affiliation{Department of Physics, Clark University, Worcester, MA 01610}

\date{\today}

\begin{abstract}
 
We investigate the relation between economic growth and equality in a modified version of the agent-based asset exchange model (AEM). The modified model is a driven system that for a range of parameter space is effectively ergodic in the limit of an infinite system. We find that the belief that ``a rising tide lifts all boats'' does not always apply, but the effect of growth on the wealth distribution depends on the nature of the growth. In particular, we find that the rate of growth, the way the growth is distributed, and the percentage of wealth exchange determine the degree of equality. We find strong numerical evidence that there is a phase transition in the modified model, and for a part of parameter space the modified AEM acts like a geometric random walk.
\end{abstract}


\maketitle

It is a wildly held belief that economic growth benefits all members of society~\cite{review, gud, jfk}. 
The purpose of this work is to test this idea in a simple model economy. Economies are complex systems which have many influences that are difficult to include in simple models~\cite{review}. However, if the question is one of a statistical nature about the economy as a whole (macroeconomics), and if we obtain the same behavior in different models, we can have some confidence that the models suggest something meaningful about real economies. 

In the original version of the asset exchange model (AEM) an equal amount of wealth is initially distributed to $N$ agents~\cite{boghos}. A pair of agents is then chosen at random and a fraction of the wealth $\alpha$ of the poorer agent is transferred from the loser of a coin flip to the winner with probability 1/2. 
The total wealth $W = \sum_{i}w_{i}(t)$ is fixed, where $w_{i}(t)$ is the wealth of the $i$th agent at time $t$~\cite{boghos, angle, krapiv}. After many time steps, the wealth is concentrated among increasingly fewer individuals, culminating as $t\rightarrow \infty$ in a single agent holding a fraction of the wealth which approaches one~\cite{boghos, angle, ispo,krapiv}. 

In this work the AEM is modified so that after $N$ exchanges (one unit of time) an amount $\mu W(t)$ is added to the system resulting in exponential growth $W(t) = W(0)e^{\mu t}$ with $\mu>0$. To distribute the growth to the agents we calculate the quantity
\be
\label{wealth-sum}
S(t) = \sum_{i}^{N} w^{\gamma}_{i}(t),
\ee
where the sum is over the $N$ agents and the parameter $\gamma \geq 0$.
The increase in wealth due to the growth $\Delta W(t) = \mu W(t)$ is assigned to the $i$th agent as
\be
\label{delta Wi}
\Delta w_{i}(t) = \mu W(t){w^{\gamma}_{i}(t)\over S(t)}.
\ee
For $\gamma = 0$ the increase in wealth generated by the growth is distributed equally. As $\gamma$ increases, the allocation of the increased wealth is weighted more heavily toward the agents with greater wealth at time $t$. Three parameters determine the wealth distribution in our modified asset exchange model (MAEM): 
$\gamma$, $\mu$, and $\alpha$. In this work we focus primarily on the effect of different values of $\gamma$. 

\begin{figure}[t]
\centering
\includegraphics[scale=0.6]{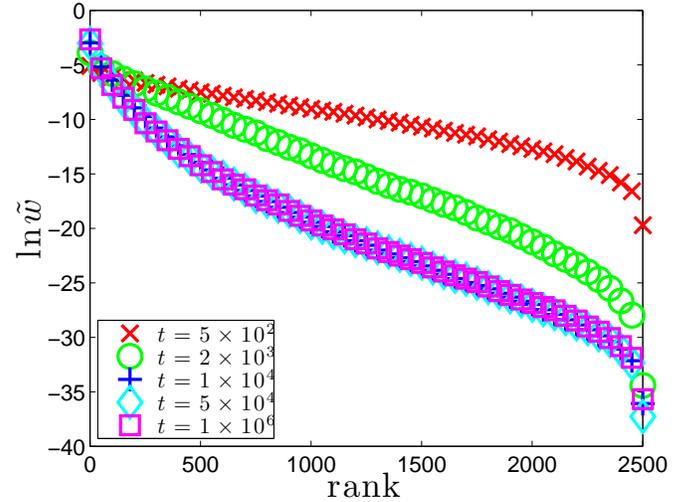}
\caption{(color online). Natural log of the rescaled wealth $\ws$ as a function of the rank of $N=2500$ agents for $\gamma=0.9$, $\mu=1\times 10^{-3}$, and $\alpha=0.1$, for different times. The rescaled wealth after $t\sim1\times 10^4$ collapses onto a single curve, indicating a rescaled steady state distribution.}
\label{fig:Fig1}
\end{figure}

In Fig.~1 we plot the distribution of the rescaled wealth $\ws = w/W(t)$ for $N = 2500$ agents as a function of the rank of their wealth for $\gamma = 0.9$, $\mu=1\times 10^{-3}$, and $\alpha=0.1$ at different times.
After an initial transient the distributions collapse onto a single curve indicating that the wealth distribution rescaled by the total wealth reaches a steady state, which we refer to as a {\it rescaled steady state}. During the transient the wealth disparity between rich and poor agents grows until the \rss\ is established. Once the \rss\ is reached, the form of the distribution (and the ratio of the wealth between rich and poor ranks) remains fixed, while the wealth in every rank grows as $e^{\mu t}$. Hence for $\gamma < 1$ a rising tide does raise all boats.

As $\gamma\rightarrow 1^{-}$ for fixed $\mu, \alpha \neq 0$, the \rss\ takes longer to establish, and the wealth distribution is less equal; i.e., the wealthiest have a greater share of the total wealth when the \rss\ is reached. 

In Fig.~2 we plot the wealth distribution as a function of rank for $\gamma = 1.1$, $\mu=1\times 10^{-3}$, and $\alpha=0.1$. No steady state is reached in the rescaled distribution. As wealth is added to the system, it is transferred to the wealthy via the exchange mechanism so inequality continues to grow. 

\begin{figure}[t]
\centering
\includegraphics[scale=0.6]{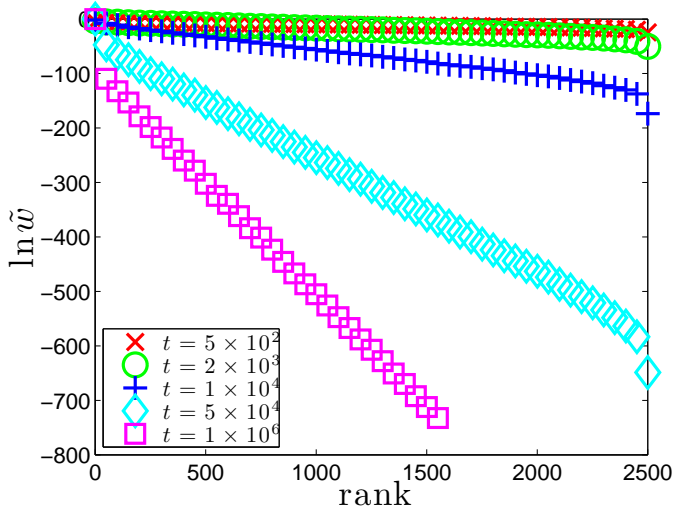}
\caption{(color online). Natural log of the rescaled wealth $\ws$ as a function of rank for $N=2500$ agents with $\gamma=1.1$, $\mu=1\times 10^{-3}$, and $\alpha=0.1$. The slope of the distribution decreases in time, and there is no rescaled steady state distribution.}
\label{fig:Fig2}
\end{figure}

For $\gamma \geq 1$ there is no \rss, and the wealth added by growth goes to the wealthiest agents while the wealth of the remaining agents declines with time. This behavior implies that there is a phase transition at $\gamma = 1$. However, the behavior of the system as a function of $\gamma$ is complicated.

\begin{figure}
\centering
\subfigure[~$\gamma=0.9$.]{\includegraphics[width = 55mm]{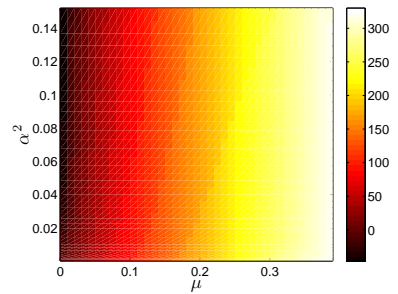}\label{fig:Fig3a}}\hspace{5mm}
\subfigure[~$\gamma=1.0$.]{\includegraphics[width = 55mm]{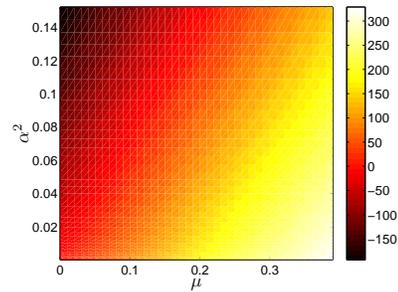}\label{fig:Fig3b}}\hspace{5mm}
\subfigure[~$\gamma=1.1$.]{\includegraphics[width = 55mm]{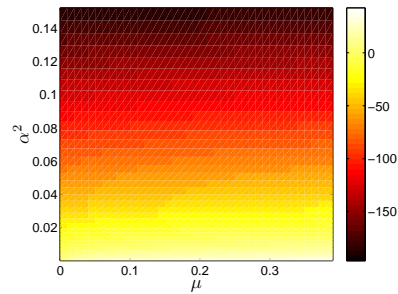}\label{fig:Fig3c}}
\caption{(color online) Natural log of the average wealth of the poorest 25 agents out of 2500 for (a) $\gamma=0.9$, (b) $\gamma=1.0$, and (c) $\gamma=1.1$. The log of the wealth is represented by shades of gray. For $\gamma<1$, the rescaled wealth of the poorest 25 agents depends mainly on $\mu$, whereas for $\gamma>1$ it mainly depends on $\alpha^2$. For $\gamma=1$, the rescaled wealth depends on both $\mu$ and $\alpha^2$.}
\label{fig:Fig3} 
\end{figure}

In Fig.~3 we plot the average wealth of the poorest 1\% of the agents (out of a total of 2500) at $t=1000$ for different values of $\gamma$, $\mu$, and $\alpha$. For $\gamma < 1$ the average wealth of the poorest agents in the \rss\ is weakly dependent on $\alpha$ but depends strongly on $\mu$. For $\gamma > 1$ there is no \rss, but the average wealth of the poorest 1\% depends strongly on $\alpha$ and depends weakly on $\mu$. For $\gamma = 1$ there is a boundary at $\Delta = \mu - k \alpha^{2}=0$ where $k\approx 1.5$. For $\Delta > 0$ the wealth of the poorest $1\%$ increases with $t$, while for $\Delta < 0$ the wealth of the poorest $1\%$ decreases with $t$. This behavior is similar to the geometric random walk~\cite{ole}, for which there is a boundary between growth and decay for the wealth of the poorer agents (members of an ensemble) at $\Delta_g = \mu-\sigma^{2}/2 = 0$; $\sigma$ is the amplitude of the random noise and, as in this work $\mu$ governs the geometric growth. Similar to the MAEM, the poorer agents' wealth in the geometric random walk grows  for $\Delta_g > 0$ and decays for $\Delta_g < 0$. 

\begin{figure}
\centering
\subfigure[~$\gamma=0.9$.]{\includegraphics[width = 55mm]{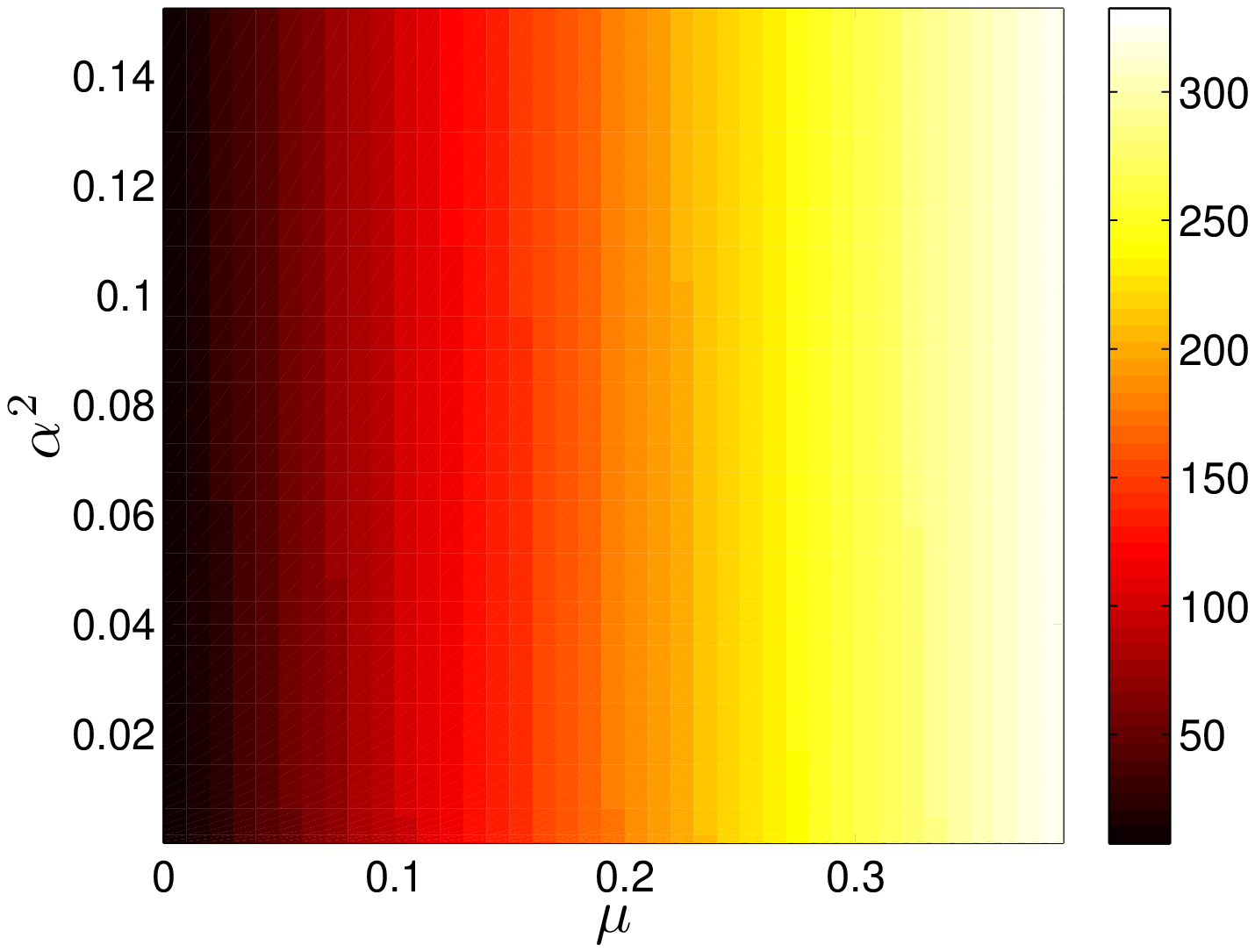}\label{fig:Fig4a}}\hspace{5mm}
\subfigure[~$\gamma=1.0$.]{\includegraphics[width = 55mm]{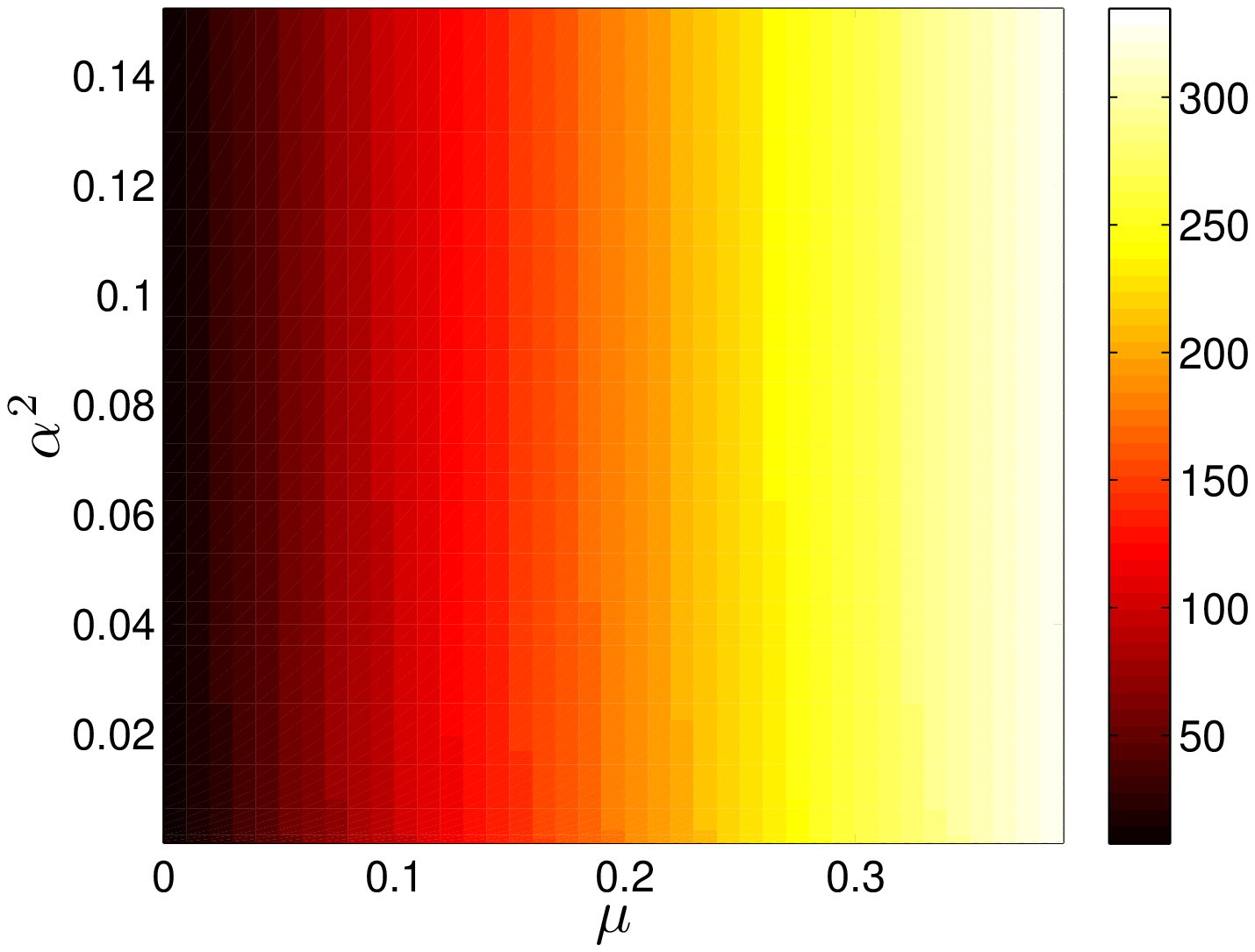}\label{fig:Fig4b}}\hspace{5mm}
\subfigure[~$\gamma=1.1$.]{\includegraphics[width = 55mm]{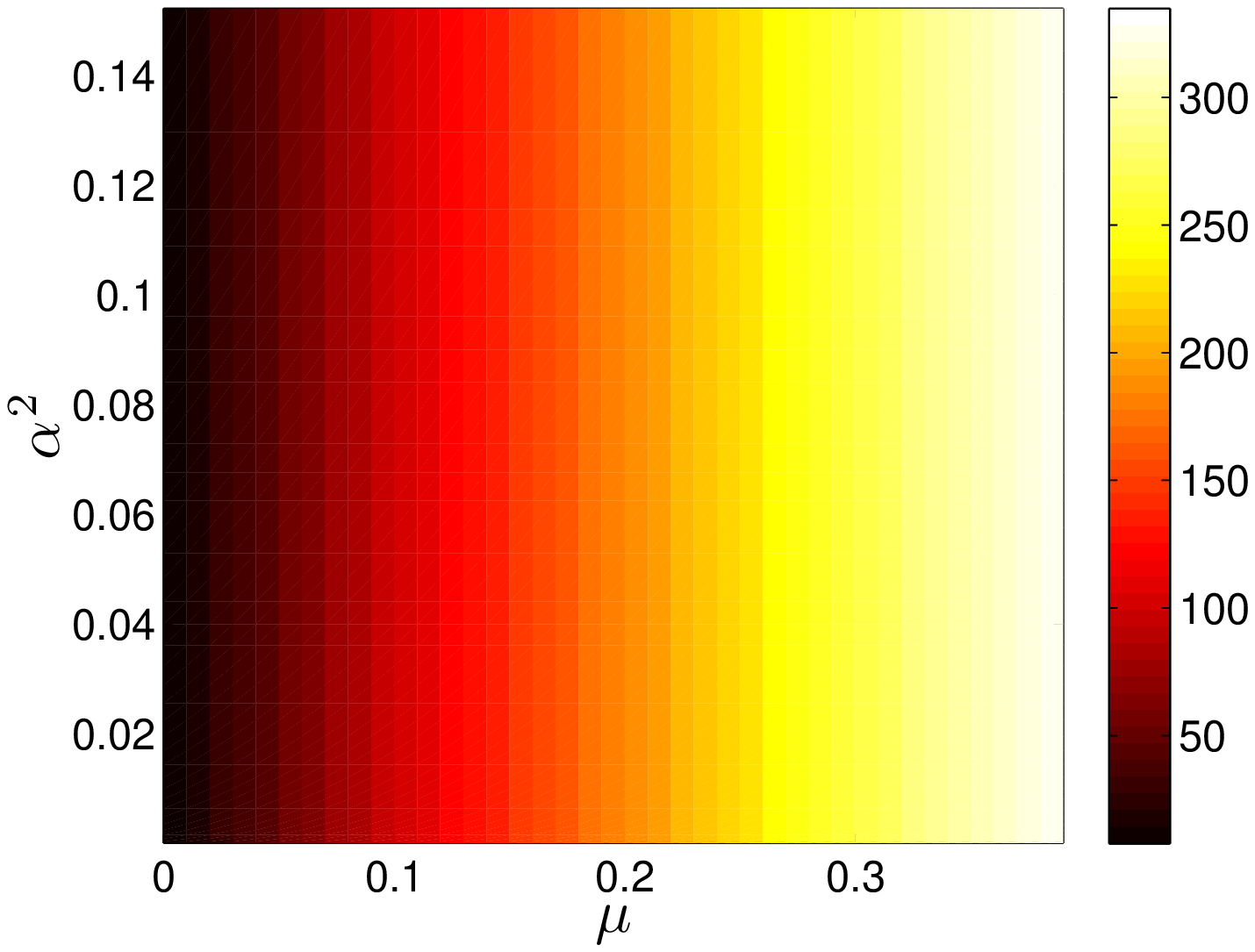}\label{fig:Fig4c}}
\vspace{-0.25cm}
\caption{(color online) Natural log of the average wealth of the richest 25 agents out of 2500 with (a) $\gamma=0.9$, (b) $1.0$, and (c) $1.1$. The log of the wealth is represented by shades of gray. For all three values of $\gamma$, the wealth depends only on $\mu$.}
\label{fig:Fig4}
\end{figure}

The behavior of the poorest $1\%$ differs from the evolution of the richest $1\%$ as can be seen by comparing Figs.~3 and 4. For the richest there is no sign of a boundary and the growth weakly depends on $\alpha$ for all $\gamma$. A phase transition at $\gamma=1$ for the wealthiest is found by looking at the rescaled  wealth of the richest agent for $t \gg 1$ as $N\rightarrow \infty$. For $\gamma < 1$ this rescaled wealth goes to zero as $N\rightarrow \infty$, and for $\gamma > 1$  it goes to a nonzero constant (see Fig.~5).

\begin{figure}[t]
\centering
\includegraphics[scale=0.6]{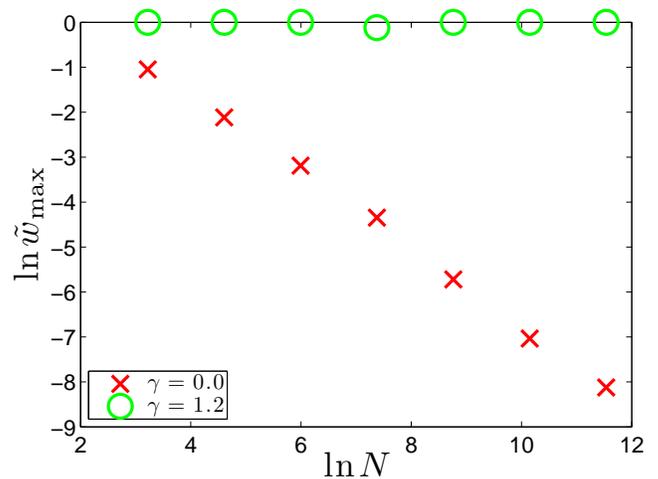}
\vspace{-0.5cm}
\caption{(color online) Plot of the natural log of the rescaled wealth of the richest agent, $\ws_{\max}$, as a function of $N$ for $\gamma < 1$ ($\times$) and $\gamma > 1$ ($\circ$). For $\gamma=0.0$, the rescaled wealth decays as $1/N$ whereas for $\gamma=1.2$, the rescaled wealth approaches a nonzero constant which is close to $1$.}
\label{fig:Fig5}
\end{figure}

An important property is economic mobility. To determine the mobility
we measure the Pearson correlation function~\cite{pearson} $C(t)$ of the rank in the steady state.
\be
\label{pear-cor}
C(t) = {\sum_{i}\big [ R_{i}(t) - {\overline R}(t)\big]\big [ R_{i}(0) - {\overline R}(0)\big ]\over \sqrt{\big [\sum_{j}\big (R_{j}(t) - {\overline R}_{j}(t)\big )^{2}\big ]\big [\sum_{k}\big (R_{k}(0) - {\overline R}_{k}(0)\big )^{2}\big ]}},
\ee
where $R_{j}(t)$ is the rank of the $j$th agent and ${\overline R}(t)=N/2$ is the ensemble average of the rank. As can be seen from Fig.~6, $C(t)\rightarrow 0$ as $t\rightarrow \infty$ for $\gamma<1$, which indicates that the rank of the agents as $t\rightarrow \infty$ is not correlated with their rank at $t = 0$. Hence, there is nonzero mobility in the MAEM for $\gamma < 1$. For $\gamma \geq 1$ $C(t)$ approaches a constant, indicating that there is a strong correlation of the rank at different times. Hence, there is a lack of mobility for $\gamma \geq 1$. 

\begin{figure}[t]
\centering
\includegraphics[scale=0.6]{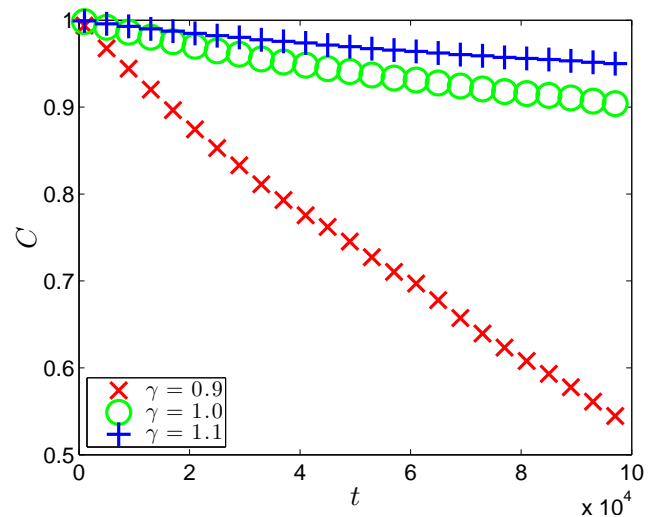}
\vspace{-0.75cm}
\caption{(color online) The Pearson correlation function as a function of $t$ for $\gamma=0.9$ ($\times$), $1.0$ ($\circ$), and $1.1$ ($+$). For $\gamma<1$, $C(t)$ decays to zero, indicating that the mobility is nonzero. In contrast, for $\gamma \geq 1$, there is a high correlation even as $t\rightarrow \infty$, which suggests a lack of mobility in the agents' ranks.}
\label{fig:Fig6}
\end{figure}

For $\gamma < 1$ the nonzero mobility suggests that the system is ergodic in the sense that the time averaged rescaled wealth of each agent equals the ensemble average of the rescaled wealth. This equality is known as effective ergodicity~\cite{tm}. To test the MAEM for effective ergodicity we use the rescaled wealth of the $i$th agent, $\ws_{i}(t)$, to define
a wealth metric as~\cite{tm}
\be
\label{metric}
\Omega(t) = {1\over N}\sum_{i=1}^{N}\big [ {\overline \ws}_{i}(t) - {<}\ws(t){>}\big]^{2},
\ee
where
\be
{\overline \ws}_{i}(t) = {1\over t}\!\int_{0}^{t} \ws_{i}(t')\,dt'
\ee
and
\be
{<}\ws(t){>} = {1\over N} \sum_{i=1}^{N}{\overline \ws}_{i}(t).
\ee

If the system is effectively ergodic, $\Omega(t)\propto 1/t$~\cite{tm}. From Fig.~7 we see that the $t$ dependence of $\Omega(0)/\Omega(t)$ indicates effective ergodicity for $\gamma <1$ and the absence of effective ergodicity for $\gamma > 1$. We conclude from the $t$-dependence of $C(t)$ and $\Omega(t)$ that the system is ergodic for $\gamma < 1$ but not for $\gamma > 1$.

\begin{figure}[t]
\centering
\includegraphics[scale=0.6]{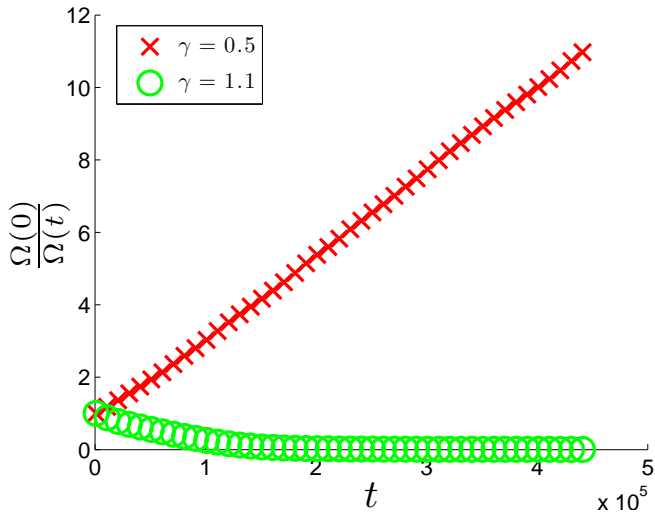}
\vspace{-0.5cm}
\caption{(color online) The inverse rescaled wealth metric as a function of $t$ for $\gamma=0.5$ ($\times$) and $\gamma = 1.1$ ($\circ$) with $\mu=1\times 10^{-4}$ and $\alpha=0.01$. For $\gamma < 1$, $\Omega(0)/\Omega(t)$ is linear indicating that the system is effectively ergodic; for $\gamma >1$ the system is non-ergodic.}
\label{fig:Fig7}
\end{figure}

Our numerical results suggest that the effect of adding growth to the AEM is to produce two ``phases.'' For $\gamma > 1$ the system does not reach a \rss, and in the limit $t\rightarrow \infty$ the wealthiest agent has a finite fraction of the total wealth as $N\rightarrow \infty$. The system is not ergodic in the sense explained above and has some of the characteristics of the geometric random walk which is not ergodic~\cite{ole,ole-bill}. For $\gamma < 1$ the system is effectively ergodic and reaches a \rss. The wealthiest agent has a zero fraction of the total wealth as $N\rightarrow \infty$. For $\gamma = 1$ the system has another transition at $\mu - k \alpha^{2} = 0$, and its behavior is similar to the geometric random walk.

The nature of these phases and the transitions is not well understood. For example, the time required for the system to reach a \rss\ as $\gamma\rightarrow 1^-$ diverges as $(1-\gamma)^{-x}$ with $x\sim 1$, suggesting a critical point~\cite{kang-et}. However, the fraction of wealth possessed by the richest agent jumps from 0 to $\sim 1$ at $\gamma = 1$. This finite discontinuity suggests that the transition is first-order.

It is very interesting that the MAEM is a driven system that might, for some range of parameter space, be treated by equilibrium methods in the limit of an infinite system.
In this way it is similar to a driven dissipative system used to model earthquake faults~\cite{rundle}. The latter model is ergodic and can be described by a Boltzmann distribution in the infinite stress transfer limit. Among the several challenges this work presents is to better understand how   exchange mechanisms in the limit $N \to \infty$ results in ergodicity in driven systems and how to identify the ``thermodynamic'' control parameters as well as quantities such as the order parameter and the nature of the phase transitions.

The economic implications of the MAEM are also quite interesting. Our results suggest that there exists a transition or tipping point. As $\gamma$ increases, the benefits of growth are weighted more toward the wealthy, and wealth inequality increases. However, as long as $\gamma < 1$, the wealth of all ranks grows at the same rate once the \rss\ is reached. If the benefits of growth are skewed too much toward the wealthy ($\gamma > 1$), the poor and middle rank agents no longer benefit from the growth, and the richest agent eventually accrues all the wealth.

There is some controversy as to whether economic systems are in equilibrium or even exhibit effective ergodicity~\cite{ole,ole-bill, yak}. As discussed in Refs.~\cite{ole,ole-bill}, the geometric random walk is not ergodic, and hence equilibrium methods do not apply. However, one of the phases of the MAEM is effectively ergodic and might be described by an equilibrium approach. Because there is no reason to believe that parameters such as $\gamma$ are temporal constants in real economies, the MAEM suggests that the applicability of equilibrium methods may be situational and vary with time.

We acknowledge useful conversations with O. Peters, J. I .Ogren and A. Gabel.

\end{document}